\documentclass[prd,showpacs]{revtex4}
\usepackage{graphicx}

\usepackage{amsmath,amsfonts,amssymb}
\newcommand{\be}{\begin{equation}}
\newcommand{\ee}{\end{equation}}
\newcommand{\bea}{\begin{eqnarray}}
\newcommand{\eea}{\end{eqnarray}}

\def\d{\rm{d}}

\def\S{\rm{S}}

\newcommand{\stg}{{\sqrt{-\tilde{g}}}}
\newcommand{\sg} {{\sqrt{-g}}}

\newcommand{\bi} {\begin{itemize}}
\newcommand{\ei} {\end{itemize}}
\newcommand{\ben} {\begin{enumerate}}
\newcommand{\een} {\end{enumerate}}

\newcommand{\la} {{\lambda}}
\newcommand{\al} {{\alpha}}
\newcommand{\ka} {{\kappa}}

\newcommand{\Om} {{\Omega}}
\newcommand{\om} {{\omega}}

\newcommand{\si} {{\sigma}}

\newcommand{\Ga} {{\Gamma}}

\def\v1{\vspace{1cm}}
\def\be{\begin{equation}}
\def\ee{\end{equation}}
\def\bc{\begin{center}}
\def\ec{\end{center}}

\newcommand{\tidd} {{\ddot{\tilde{a}}}}
\newcommand{\tid} {{\dot{\tilde{a}}}}
\newcommand{\tia} {{\tilde{a}}}
\newcommand{\tit} {{\tilde{t}}}

\begin{document}

\begin{flushright}
Preprint MPG-VT-UR 262/05
\end{flushright}

%\baselineskip18pt
%\title{Puzzles of isotropic and anisotropic conformal cosmologies.
\title{The conformal status of $\om = -3/2$ Brans-Dicke cosmology.}

\author{Mariusz P. D\c{a}browski}
\email{mpdabfz@sus.univ.szczecin.pl}
\affiliation{\it Institute of Physics, University of Szczecin, Wielkopolska 15,
          70-451 Szczecin, Poland}

\author{Tomasz Denkiewicz}
\email{tomasz.denkiewicz@uni-rostock.de}
\affiliation{\it Fachbereich Physik, Universit\"at Rostock,
Universit\"atsplatz 3, D-18051 Rostock, Germany}
\affiliation{\it Institute of Physics, University of Szczecin, Wielkopolska 15,
          70-451 Szczecin, Poland}

\author{David Blaschke}
\email{david.blaschke@physik.uni-rostock.de}
\affiliation{\it Fachbereich Physik, Universit\"at Rostock,
Universit\"atsplatz 3, D-18051 Rostock, Germany}
\affiliation{\it Gesellschaft f\"ur Schwerionenforschung (GSI) mbH, Planckstrasse 1,
D-64291 Darmstadt, Germany}
\affiliation{\it Bogoliubov Laboratory for Theoretical Physics,
Joint Institute for Nuclear Research, RU-141980 Dubna, Russia}

\date{\today}

\begin{abstract}

Following recent fit of supernovae data to Brans-Dicke theory which favours the model
with $\omega = - 3/2$ \cite{fabris} we discuss the status of this special case of
Brans-Dicke cosmology in both isotropic and anisotropic framework.
It emerges that the limit $\omega = -3/2$ is consistent only with
the vacuum field equations and it makes such a Brans-Dicke theory
conformally invariant. Then it is an example of the conformal relativity
theory which allows the invariance with respect to conformal transformations of the metric.
Besides, Brans-Dicke theory with $\omega = -3/2$ gives a border between
a standard scalar field model and a ghost/phantom model.

In this paper we show that in $\omega = -3/2$ Brans-Dicke theory, i.e., in the
conformal relativity there are no isotropic Friedmann solutions of non-zero
spatial curvature except for $k=-1$ case. Further we show that this $k=-1$
case, after the conformal transformation into the Einstein frame, is just the Milne
universe and, as such, it is equivalent to Minkowski spacetime. It generally means that
only flat models are fully consistent with the field equations. On the other hand, it is
shown explicitly that the anisotropic non-zero spatial curvature
models of Kantowski-Sachs type are admissible in $\omega = -3/2$ Brans-Dicke theory.
It then seems that an additional scale factor which
appears in anisotropic models gives an extra deegre of freedom and makes it less
restrictive than in an isotropic Friedmann case.

\end{abstract}


\maketitle

\section{Introduction}
\setcounter{equation}{0}

The fundamental equations of physics
such as Maxwell equations, massless Dirac equation and massless Klein-Gordon
equation are invariant with respect to conformal transformations of the metric
\cite{birell}. However, the Einstein equations are not invariant with respect to these
transformations. Modifications of these equations which involve
the {\it conformal coupling} of a metric to a scalar field which leads to
conformal invariance are called conformal relativity and the examples of
such a theory have been studied \cite{HN,canuto,chern,be74,narlikar,penrose,deser70}.
Recently, in one of the versions of these theories the geometrical evolution
of the universe was reinterpreted as an evolution of
mass represented by a scalar field in a flat universe \cite{bbpp,iap,guendelman,relatnn,frampton}.
The idea is quite interesting and it may help to
resolve the problem of the dark energy in the universe
\cite{weinberg_kosmkonst,supernovae,ptw}. Similar ideas
have been developed in yet another modification of general
relativity called Self Creation Cosmology \cite{barber} in which
the dark energy problem together with a series of other cosmological problems including
Pioneer spacecraft puzzle \cite{pioneer} have been studied. In yet
another proposal, a conformal transformation helps to suppress
the cosmological constant in the conformal frame \cite{jackiw}.

The simplest example of a conformally invariant theory is the
Brans-Dicke theory with Brans-Dicke parameter $\omega = - 3/2$ \cite{relatnn}.
Amazingly, the recent fit to supernovae data \cite{supernovae} shows that, despite
local gravitational tests which give the constraint $\omega > 1000$,
supernovae favour exactly the value of $\omega = - 3/2$ \cite{fabris}.
It is also interesting to note that, apart from all the above, the Brans-Dicke theory
\cite{bd} with $\omega = -3/2$ gives a border line between
standard scalar field models and ghost/phantom models in the Einstein frame \cite{phantom}.

In this paper we discuss the exact conformal relativistic solutions, i.e., $\omega =-3/2$
Brans-Dicke solutions of isotropic Friedmann, anisotropic Kantowski-Sachs, axisymmetric
Bianchi I, and Bianchi III type. We derive them from a more general context of
Brans-Dicke theory taking the limit $\omega =-3/2$ and compare them
with analogous solutions in low-energy superstring cosmology
\cite{polchinsky,bd97,superjim,dab5}.

The paper is organized as follows. In Section II we discuss the
status of $\omega=-3/2$ Brans-Dicke theory as conformal
relativity. In Section III we present isotropic conformal
cosmology solutions in the Jordan frame and then in the Einstein frame.
We apply various time
parametrizations in order to look for the one which is
non-singular in the $\omega=-3/2$ limit. In Section IV we find
anisotropic conformal cosmology solutions of Kantowski-Sachs type
in the Jordan frame and present them directly in terms of the
cosmic time coordinate instead of the parametric time as it was
given in the previous literature. In Section V we give our
conclusions.

\section{Conformal relativity as $\omega=-3/2$ Brans-Dicke theory}

\setcounter{equation}{0}

Suppose that we have two spacetime manifolds ${\cal M}, \tilde{\cal M}$ with
metrics $g_{\mu\nu}, \tilde{g}_{\mu\nu}$ and {\it the same} coordinates
$x^{\mu}$. We say that the two manifols are {\it conformal} to
each other if they are related by the following {\it conformal
transformation}
\bea
\label{conf_trafo}
\tilde{g}_{\mu\nu} &=& \Omega^2(x) g_{\mu\nu}~,
\eea
and $\Omega(x)$ which is called a conformal factor must
be a twice-differentiable function of coordinates $x^{\mu}$ and
lie in the range $0<\Omega<\infty$.
The conformal transformations shrink or stretch the distances between the two points
described by the same coordinate system $x^{\mu}$ on the manifolds
${\cal M}, \tilde{\cal M}$, respectively, but they preserve the angles between vectors
(in particular, between null vectors which define the light cones) which
leads to a conservation of the (global) causal structure of the manifold \cite{hawk_ellis}.
This means that null geodesics are left intact while the timelike
geodesics are modified by conformal transformations.
If we take $\Om =$ const. we deal with the so-called {\it scale
transformations} \cite{maeda}. In fact, conformal transformations are
{\it localized} scale transformations $\Om = \Om(x)$.

On the other hand, the {\it coordinate transformations} $x^{\mu} \to \tilde{x}^{\mu}$ only
relabel the coordinates and do not change geometry and
they are entirely {\it different} from conformal transformations \cite{narlikar}.
This is crucial since conformal
transformations lead to a {\it different physics} on
conformally related manifolds ${\cal M}, \tilde{\cal M}$ \cite{maeda}.
Since this will usually be related to a {\it different coupling}
of a physical field to gravity we will be talking about different
{\it frames} in which the physics is studied (see also Ref. \cite{flanagan} for
a slightly different view).

In this paper we discuss the following conformally invariant action
\cite{chern,penrose,relatnn}
\bea
\label{tildeconfinvpi}
\tilde{\S} &=&
 \frac{1}{16\pi}\frac{1}{2}\int~\d^4x\stg\tilde{\Phi}\left (
  \frac{1}{6}\tilde{R}\tilde{\Phi} -\stackrel{\sim}{\Box}\tilde{\Phi}\right
  )~.
\eea
It composes of the scalar field (the dilaton) which is conformally
coupled to the metric (the graviton). Conformal invariance means that the application of
the conformal transformation (\ref{conf_trafo}) with the conformal
factor chosen to be
\bea
\label{PhitildetoPhi}
\tilde{\Phi} &=& \Om^{-1} \Phi~,
\eea
brings (\ref{tildeconfinvpi}) to the same form, i.e.,
\bea
\label{confinvpi}
\S &=&
 \frac{1}{16\pi}
\frac{1}{2}\int~\d^4x\sg{\Phi}\left (\frac{1}{6}{R}{\Phi}-{\Box}\Phi \right
 )~,
\eea
where all the quantities have no tildes.

We do not admit any matter part into the action (or the matter
energy-momentum tensor into the field equations) so that, despite
we have a dilaton field, we formally deal with vacuum field
equations. However, they do not look like vacuum field equations since they
are obtained as a result of a non-minimal coupling of the dilaton
to the graviton.

These actions (\ref{tildeconfinvpi}) and (\ref{confinvpi}) are
usually represented in a different form by the application of the expression
for a covariant d'Alambertian for a scalar field in general relativity
\bea
\label{boxpartial}
\stackrel{\sim}{\Box}\tilde{\Phi} &=& \frac{1}{\sqrt{-\tilde{g}}}
\stackrel{\sim}{\partial}_{\mu} \left(\sqrt{-\tilde{g}}
\stackrel{\sim}{\partial}^{\mu}{\tilde{\Phi}}
\right)~,
\eea
which after integrating out the boundary term, gives \cite{maeda}
\bea
\label{boundarytilde}
\tilde{S} &=& \frac{1}{16\pi} \frac{1}{2} \int d^4x \sqrt{-\tilde{g}} \left[
\frac{1}{6} \tilde{R} \tilde{\Phi}^2 +
\stackrel{\sim}{\partial}_{\mu}\tilde{\Phi}\stackrel{\sim}{\partial}^{\mu}\tilde{\Phi}
\right]~,
\eea
and the second term is just a kinetic
term for the scalar field (cf. \cite{birell,hawk_ellis}). The
equations (\ref{boundarytilde}) are of course also conformally invariant, since
the application of the formulas (\ref{det}), (\ref{ricciscalar4})
and (\ref{PhitildetoPhi}) together with the appropriate
integration of the boundary term gives the same form of the equations
\bea
\label{boundary}
S &=& \frac{1}{16\pi} \frac{1}{2} \int d^4x \sqrt{-g} \left[
\frac{1}{6} R \Phi^2 +
{\partial}_{\mu}\Phi {\partial}^{\mu}{\Phi}
\right]~.
\eea
In fact, due to the type of non-minimal coupling of gravity to a scalar
field $\tilde{\Phi}$ or $\Phi$ in (\ref{boundarytilde}) and
(\ref{boundary}) and the relation to Brans-Dicke theory
 we say that these equations are presented in the {\it
Jordan frame} \cite{jordan,maeda}.

It is worth noticing that adding the self-interacting scalar field potential
\bea
\label{selfint}
\tilde{U}(\tilde{\Phi})=\frac{\tilde{\la}}{4}\tilde{\Phi}^4~,
\eea
with the coupling
constant $\tilde{\la}$ is conformally-invariant (but only in $D=4$
spacetime dimensions \cite{relatnn}). In order to see this, we start with the
action with self-interaction potential which under conformal transformation
changes as
\bea
\label{selfinteraction}
\S &=&
\frac{1}{16\pi} \frac{1}{2}\int~\d^4x~\stg\left[
\frac{1}{6} \tilde{R}\tilde{\Phi}^2+
  \stackrel{\sim}{\partial}_{\mu}\tilde{\Phi}\stackrel{\sim}{\partial}^{\mu}\tilde{\Phi}
+\frac{\tilde{\la}}{4}\tilde{\Phi}^4 \right
]\nonumber  \\
&=&
\frac{1}{16\pi} \frac{1}{2}\int~\d^4x~\sg\left [
\frac{1}{6} R \Phi^2 +
{\partial}_{\mu}\Phi {\partial}^{\mu}{\Phi}
+  \frac{\tilde{\la}}{4}\Phi^4\right ] ~,
\eea
so that the new self-interaction potential reads as
\bea
U(\Phi)=\frac{\tilde{\la}}{4}\Phi^4~.
\eea
This fact was used in Ref. \cite{jackiw} where in one of the
frames the cosmological constant related to Anti-deSitter solution
was suppressed due to the quantum arguments in the flat Minkowski
second frame.

The conformally invariant actions (\ref{tildeconfinvpi}) and (\ref{confinvpi})
are the basis to derive the equations of motion via the variational principle.
The equations of motion for scalar fields $\tilde{\Phi}$ and $\Phi$
are conformally invariant
\bea
\label{eom_1}
\left
  (\stackrel{\sim}{\Box}-\frac{1}{6} \tilde{R} \right ) \tilde{\Phi} =
  \Om^{-3} \left(\Box - \frac{1}{6} R\right) \Phi &=& 0~,
\eea
and they have the structure of the Klein-Gordon equation with the
mass term replaced by the curvature term ~\cite{chern}. In fact,
this leads to a violation of the strong equivalence principle which may
either be constrained by observations or admitted in the very
early universe. The conformally invariant Einstein equations are obtained from
variation of $\tilde{S}$ with respect to the metric $\tilde{g}_{\mu\nu}$
and read as
\bea
\label{eom3}
\left( \tilde{R}_{\mu\nu}- \frac{1}{2}\tilde{g}_{\mu\nu}\tilde{R}
\right) \frac{1}{6} \tilde{\Phi}^2 + \frac{1}{6}
\left[ 4 \tilde{\Phi}_{,\mu}\tilde{\Phi}_{,\nu} -
\tilde{g}_{\mu\nu}
\tilde{\Phi}_{,\al}\tilde{\Phi}^{,\al} \right] + \frac{1}{3} \left[ \tilde{g}_{\mu\nu}
\tilde{\Phi} \stackrel{\sim}{\Box} \tilde{\Phi} -
\tilde{\Phi} \tilde{\Phi}_{\tilde{;}\mu\nu} \right] &=& 0~.
\eea

Applying (\ref{riccitensor1}), (\ref{ricciscalar4}), (\ref{PhitildetoPhi})
and (\ref{covdertilde}) into
(\ref{eom3}) gives the same {\it conformally invariant} form of
the field equations as
\bea
\label{eom4}
\left( R_{\mu\nu}- \frac{1}{2}g_{\mu\nu}R
\right) \frac{1}{6} \Phi^2 + \frac{1}{6}
\left[ 4 \Phi_{,\mu}\Phi_{,\nu} -
g_{\mu\nu}
\Phi_{,\al} \Phi^{,\al} \right] + \frac{1}{3} \left[ g_{\mu\nu}
\Phi \Box \Phi -
\Phi \Phi_{;\mu\nu} \right] &=& 0~.
\eea
These are exactly the same field equations as in the Hoyle-Narlikar theory \cite{narlikar}
(see also Canuto-Hsieh thoery \cite{canuto}).
Note that the scalar field equations of motion (\ref{eom_1}) can be obtained by
an appropriate contraction of the equations (\ref{eom3}) and (\ref{eom4}),
so that they are not independent and do not supply any additional
information \cite{relatnn}.

We can easily relate conformal relativity to Brans-Dicke theory
using conformally invariant actions (\ref{tildeconfinvpi}) and (\ref{confinvpi})
in the Jordan frame by defining new scalar fields $\phi, \tilde{\phi}$
as
\bea
\label{Phitophi}
\frac{1}{12} \Phi^2 = e^{-\phi}~, \hspace{0.3cm}
\frac{1}{12} \tilde{\Phi}^2 = e^{-\tilde{\phi}}~, \hspace{0.3cm}
e^{-\tilde{\phi}/2} = \Omega^{-1} e^{-\phi/2}~,
\eea
which gives these conformally invariant actions in the form
\bea
\label{eff}
S &=& \frac{1}{16\pi} \int d^4x \sqrt{-g} e^{-\phi}\left[
R + \frac{3}{2}
{\partial}_{\mu}\phi {\partial}^{\mu}{\phi}
\right]~,
\eea
\bea
\label{tildeeff}
\tilde{S} &=& \frac{1}{16\pi} \int d^4x \sqrt{-\tilde{g}} e^{-\tilde{\phi}} \left[
\tilde{R}  + \frac{3}{2}
\stackrel{\sim}{\partial}_{\mu}\tilde{\phi}\stackrel{\sim}{\partial}^{\mu}\tilde{\phi}
\right]~.
\eea
These actions, however, are special cases of the Brans-Dicke
action written down in terms of the scalar field $\phi$, i.e.,
\bea
\label{ephi}
\Phi_{BD} &=& e^{-\phi}
\eea
where $\Phi_{BD}$ is the Brans-Dicke field \cite{bd,relatnn}.
This Brans-Dicke action which is {\it not} conformally invariant reads as
\bea
\label{effom}
S &=&  \frac{1}{16\pi} \int d^4x \sqrt{-g} e^{-\phi}\left[
R  - \om
{\partial}_{\mu}\phi {\partial}^{\mu}{\phi}
\right]~
\eea
which, in view of the equations (\ref{eff}) and (\ref{effom}),
shows the equivalence of the vacuum conformal relativity with Brans-Dicke theory
provided that the Brans-Dicke parameter
\bea
\label{32}
\om &=& - \frac{3}{2}~.
\eea
On the other hand, if one takes
\bea
\om &=& - 1
\eea
in (\ref{effom}), then one obtains the low-energy-effective
superstring action (which is also {\it not} conformally invariant)
for only graviton and dilaton in the spectrum \cite{superjim,dab5}
\bea
\label{effstr}
S &=& \frac{1}{16\pi} \int d^4x \sqrt{-g} e^{-\phi}\left[
R  + {\partial}_{\mu}\phi {\partial}^{\mu}{\phi}
\right]~.
\eea

In fact, the action (\ref{effom}) represents Brans-Dicke theory in
a special frame which is known as {\it string frame} or Jordan frame. It is
because in superstring theory the coupling constant $g_s$ is related to
the vacuum expectation value of the dilaton by \cite{dab5}
\bea
\label{gs}
g_s \propto e^{\phi/2}.
\eea

The field equations which are obtained by the variation of
(\ref{effom}) with respect to the dilaton $\phi$ and the graviton
$g_{\mu\nu}$, respectively, are \cite{clw1}
\bea
\label{stringfe1}
R  + \om {\partial}_{\mu}\phi {\partial}^{\mu}{\phi} - 2\om \Box
\phi &=& 0~,\\
\label{stringfe2}
R_{\mu\nu} - \frac{1}{2} g_{\mu\nu} R &=&
(\om+1) {\partial}_{\mu}\phi {\partial}_{\nu}{\phi} -
\left( \frac{\om}{2} + 1 \right) g_{\mu\nu} {\partial}_{\rho}\phi {\partial}^{\rho}{\phi}
+ g_{\mu\nu} \Box \phi - \phi_{;\mu\nu}~.
\eea

It is interesting to note that the Ricci tensor which can be
calculated from (\ref{stringfe1})-(\ref{stringfe2}) as
\bea
R_{\mu\nu} &=&  - \phi_{;\mu\nu} + (\om + 1) \left(
{\partial}_{\mu}\phi {\partial}_{\nu}{\phi} -
g_{\mu\nu} {\partial}_{\rho}\phi {\partial}^{\rho}{\phi}
+ g_{\mu\nu} \Box \phi \right)~,
\eea
and for low-energy-effective superstring theory $\om = -1$, the whole lot of
its terms vanish. However, this is not the case in conformal
relativity $\om = -(3/2)$, for which this expression is not so simple.

Further on, we will look for the graviton-dilaton solutions of the most general
Brans-Dicke action (\ref{effom}) whose field equations are given
by \cite{clw1,bd97,superjim}
\bea
\label{pbb1}
R  + \om {\partial}_{\mu}\phi {\partial}^{\mu}{\phi} - 2\om \Box
\phi &=& 0~,\\
\label{pbb2}
R_{\mu\nu} + \phi_{;\mu\nu} - (\om + 1) \left(
{\partial}_{\mu}\phi {\partial}_{\nu}{\phi} -
g_{\mu\nu} {\partial}_{\rho}\phi {\partial}^{\rho}{\phi}
+ g_{\mu\nu} \Box \phi \right) &=& 0~,
\eea
{\it taking the conformal relativity limit} $\om = -3/2$ in the end. The
Kantowski-Sachs type of solutions of (\ref{pbb1}) and (\ref{pbb2}) for the common sector
of superstring theories (including the axion field) were given in
Ref. \cite{bd97}.

\section{Conformal Friedmann cosmology}

\setcounter{equation}{0}

We discuss Friedmann cosmology in the two conformally related frames
as given in (\ref{conf_trafo}), i.e.,
\bea
\label{FRWtildemetric}
d\tilde{s}^2 &=&  - d\tit^2 + \tia^2 \left(\frac{dr^2}{1 - kr^2} + r^2
d\theta^2 + r^2\sin^2{\theta} d\varphi^2 \right)~\\
\label{FRWmetric}
ds^2 &=& - dt^2 + a^2 \left(\frac{dr^2}{1 - kr^2} + r^2
d\theta^2 + r^2\sin^2{\theta} d\varphi^2\right)~,
\eea
and $k=0,\pm1$ is the spatial curvature index.
From (\ref{conf_trafo}), (\ref{FRWtildemetric}) and
(\ref{FRWmetric}) one can easily see that the time coordinates and
scale factors are related by \cite{clw1,bd97,polarski}
\bea
\label{cf1}
d\tit &=& \Omega dt~,\\
\label{cf2}
\tia &=& \Omega a~,
\eea
where for the full conformal invariance one has to apply the
definition of conformal factor (\ref{PhitildetoPhi}). Note that
there is a sign choice freedom in the equations
(\ref{cf1})-(\ref{cf2}) as a consequence of the conformal
equaivalence of the two metrics (\ref{FRWtildemetric}) and
(\ref{FRWmetric}).

\begin{figure}[h!]\begin{center}
\includegraphics[width=12cm]{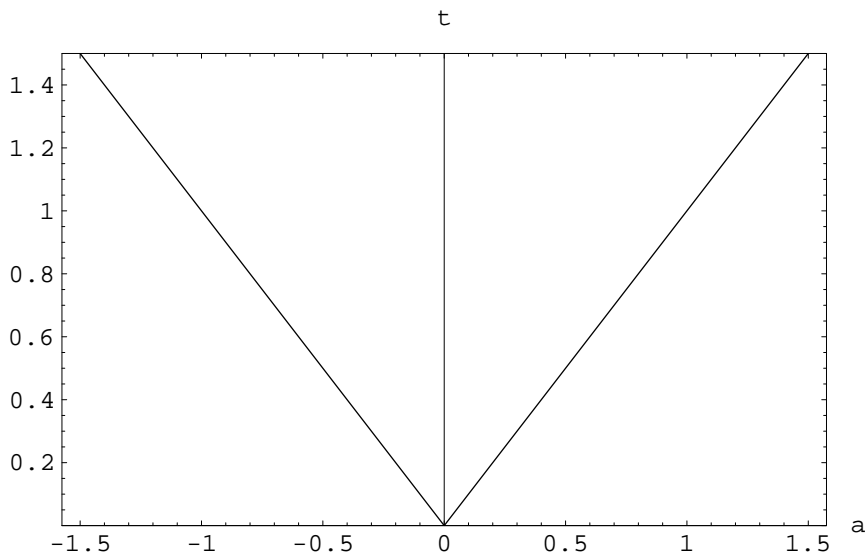}
\caption{The scale factor $a(t)$ (\ref{cgrs1}) for an isotropic conformal relativity $(\om = -3/2)$ model.
There is a curvature singularity (Big-Bang) at $t=0$.}
\label{figabst}\end{center}
\end{figure}

\begin{figure}[h!]\begin{center}
\includegraphics[width=12cm]{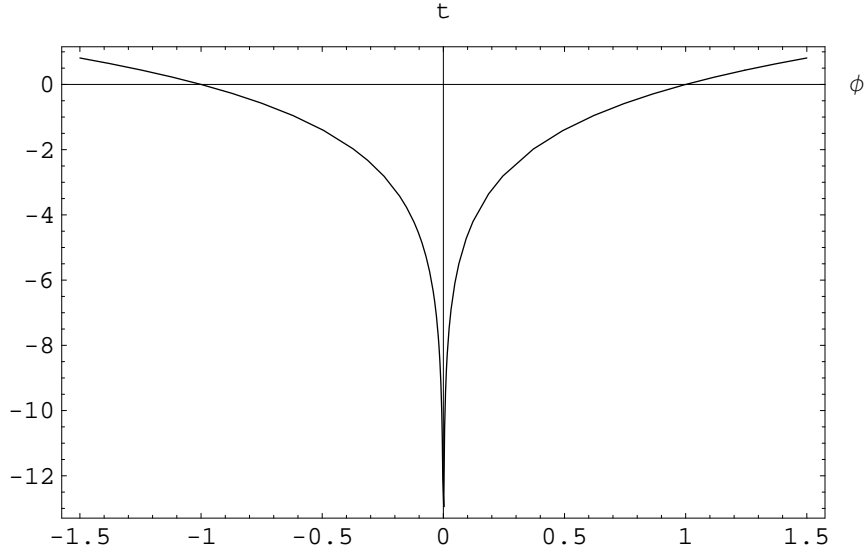}
\caption{The evolution of the scalar field $\phi(t)$ (\ref{cgrs2}) in an isotropic conformal relativity
$(\om = -3/2)$ model. There is no strong coupling singularity, since the string coupling
constant $g_s = \exp{\phi/2} \to 0$ for $t \to 0$.}
\label{figlnt}\end{center}
\end{figure}

In the string frame we use the Friedmann metric (\ref{FRWmetric})
which imposed into the equations (\ref{pbb1})-(\ref{pbb2}) for an arbitrary value of the
parameter $\om$ gives the following set of equations
\bea
\label{pbbom1}
\dot{\phi} - 3 \frac{\dot{a}}{a} &=& \frac{\ddot{\phi}}{\dot{\phi}}~,\\
\label{pbbom2}
- 3 \frac{\dot{a}^2 + k}{a^2} &=& - \left( \frac{\om}{2} + 1
\right) \dot{\phi}^2 + \ddot{\phi}~,\\
\label{pbbom3}
- 2 \frac{\ddot{a}}{a} - \frac{\dot{a}^2 + k}{a^2} &=&
\frac{\om}{2} \dot{\phi}^2 + \frac{\dot{a}}{a} \dot{\phi}~.
\eea
These equations (\ref{pbbom1})-(\ref{pbbom3}) for the flat $k=0$ Friedmann metric
give the following solutions
\bea
\label{oms1}
a(t) &=& \mid t \mid^{\frac{3(\om+1) \pm
\sqrt{3(2\om+3)}}{3(3\om+4)}}~,\\
\label{oms2}
\phi(t) &=& \frac{-1 \pm \sqrt{3(2\om+3)}}{3\om+4} \ln{\mid t
\mid}~,
\eea
where following pre-big-bang/ekpyrotic scenario
\cite{khoury01,steintur02,turok1,turok2,turok0} the solutions for
negative times are also admitted. From (\ref{pbbs1})-(\ref{pbbs2})
one can first find the pre-big-bang solutions for $\om=-1$
\cite{dab5} which are very well-known and read
\bea
\label{pbbs1}
a(t) &=& \mid t \mid^{\pm \frac{1}{\sqrt{3}}}~,\\
\label{pbbs2}
\phi(t) &=& (\pm \sqrt{3} - 1) \ln{\mid t \mid}~.
\eea
However, the conformal relativity solutions for $\om = -\frac{3}{2}$
are (see also \cite{kim})
\bea
\label{cgrs1}
a(t) &=& \mid t \mid~,\\
\label{cgrs2}
\phi(t) &=& 2 \ln{\mid t \mid}~,
\eea
and show that they do not allow for two branches $`+'$ and $`-'$
(see Figs. \ref{figabst} and \ref{figlnt}) and so they do not allow the scale
factor duality \cite{superjim}
\bea
a(t) &\to& \frac{1}{a(-t)}, \hspace{0.4cm} \phi \to \phi - 6 \ln{a}~,
\eea
which is a cosmological consequence of string duality symmetries \cite{polchinsky}.
However, unlike pre-big-bang solutions (\ref{pbbs1})-(\ref{pbbs2})
which must be regularized at Big-Bang singularity because both the curvature and the
string coupling (\ref{gs}) diverge there, the solutions (\ref{cgrs1})-(\ref{cgrs2})
do not lead to strong coupling singularity in the sense of string
theory, since the string coupling constant
\bea
g_s &=& e^{\frac{\phi}{2}} = \mid t \mid~,
\eea
is regular for $t=0$. This has an interesting analogy with the
ekpyrotic/cyclic universe scenario where, in fact, the transition through
Big-Bang singularity takes place in the weak coupling regime
\cite{turok0}.

Now let us discuss the isotropic Friedmann $k=\pm 1$ solutions of
the system (\ref{pbbom1})-(\ref{pbbom3}). After introducing a new time parameter
\bea
\zeta &=& \int \frac{dt}{a(t)},
\eea
the solutions for $k=+1$ are \cite{barrow01}
\bea
\label{confkp}
a(\zeta) &=&
\left(\sin{\zeta}\right)^{\frac{1-\sigma}{2}}\left(\cos{\zeta}\right)^{\frac{1+\sigma}{2}},~\\
\phi(\zeta) &=& - \sigma\ln(\tan{\zeta}),
\eea
while the solutions for $k=-1$ are
\bea
\label{confkn}
a(\zeta) &=&
\left(\sinh{\zeta}\right)^{\frac{1-\sigma}{2}}\left(\cosh{\zeta}\right)^{\frac{1+\sigma}{2}},~\\
\phi(\zeta) &=& - \sigma\ln(\tanh{\zeta}),
\eea
where
\bea
\label{sigma}
\sigma &=& \pm \frac{3}{3+2\omega}~.
\eea
From the definition of the parameter $\sigma$ in (\ref{sigma}) one can
see that in the conformal relativistic limit $\omega = -3/2$ this
parameter diverges, i.e., $\sigma \to \pm \infty$ and consequently
the solutions (\ref{confkp}) and (\ref{confkn}) are inappropriate.
Apparently, the status of the $\om = -3/2$ case has not been fully
cleared out so far. In particular, this case was never expressed
in terms of cosmic time instead of parametric (generalized
conformal) time, although it was presumably solved in earlier references
\cite{tupper,petzold83a,petzold83b}.

In this context we will then discuss the conformal
relativistic solutions of the system (\ref{pbbom1})-(\ref{pbbom3})
for $\omega=-3/2$ (see also \cite{kim})
\bea
\label{cc1}
\dot{\phi} - 3 \frac{\dot{a}}{a} &=& \frac{\ddot{\phi}}{\dot{\phi}}~,\\
\label{cc2}
- 3 \frac{\dot{a}^2 + k}{a^2} &=& - \frac{1}{4} \dot{\phi}^2 + \ddot{\phi}~,\\
\label{cc3}
- 2 \frac{\ddot{a}}{a} - \frac{\dot{a}^2 + k}{a^2} &=&
-\frac{3}{4} \dot{\phi}^2 + \frac{\dot{a}}{a} \dot{\phi}~.
\eea
Using a new time coordinate \cite{bd97}
\bea
\label{ttau}
dt &=& a^3 e^{-\phi} d\tau~,
\eea
the equation (\ref{cc1}) reads as
\bea
\label{cc1t}
\phi_{,\tau\tau} &=& 0~,
\eea
where $(\ldots)_{,\tau}$ describes a derivative with respect to $\tau$.
The equations (\ref{cc2}) and (\ref{cc3}) now are
\bea
\label{cc2t}
- 3 \frac{a_{,\tau}^2}{a^2} - 3 ka^4 e^{-2\phi} =
\frac{3}{4} \phi_{,\tau}^2 - 3 \frac{a_{,\tau}}{a} \phi_{,\tau} +
\phi_{,\tau\tau}~,\\
\label{cc3t}
-2 \frac{a_{,\tau\tau}}{a} + 5 \frac{a_{,\tau}^2}{a^2} - 3
\frac{a_{,\tau}}{a} \phi_{,\tau} - k a^4 e^{-2\phi} = -
\frac{3}{4} \phi_{,\tau}^2~.
\eea
The sum of (\ref{cc2t}) and (\ref{cc3t}) gives
\bea
(\ln{a})_{,\tau\tau} + 2ka^4 e^{-2\phi} &=& 0~.
\eea
or
\bea
\label{M}
(\ln{M})_{,\tau\tau} + 8kM &=& 0~,
\eea
where
\bea
\label{Mtau}
M(\tau) &=& a^4 e^{-2\phi}~.
\eea
The solution of the equation (\ref{M}) reads as
\bea
\label{Msol}
\frac{1}{\sqrt{M(\tau)}} &=& \cosh{\beta \tau} +
\sqrt{1 - \frac{4k}{\beta^2}} \sinh{\beta \tau}~,
\eea
where $\beta=$ const. and $\beta^2/4 > k$ which suggests that the only possibility
in order not to restrict the values of $\beta$ is to admit $k=-1$.
Note that (\ref{Msol}) can also be expressed as
\bea
\label{Msollim}
\frac{1}{\sqrt{M(\tau)}} &=& \cosh{\beta \tau} +
\frac{\sinh{\beta \tau}}{\mid \beta \mid \tau} \tau \sqrt{\beta^2 - 4k}~.
\eea

From (\ref{cc1t}) we have that
\bea
\phi(\tau) &=& \alpha \tau + \gamma~,
\eea
and without a loss of generality taking $\gamma=0$ we have
\bea
\label{atau}
a(\tau) &=& \frac{e^{\frac{\alpha}{2}\tau}}
{\left[\cosh{\beta \tau} +
\sqrt{1 - \frac{4k}{\beta^2}} \sinh{\beta \tau} \right]^{\frac{1}{2}}}~.
\eea
In order to deparametrize the solution (\ref{atau}) one should use
(\ref{ttau}), i.e.,
\bea
t(\tau) &=& \int{M^{3/4}e^{\phi/2}d\tau} =
\int{\frac{e^{\frac{\alpha\tau}{2}}}
{\left(\cosh{\beta \tau} + \sqrt{1 - \frac{4k}{\beta^2}} \sinh{\beta \tau}
\right)^{\frac{3}{2}}}d\tau}~.
\eea
Notice that by using the definition (\ref{Mtau}) the equations
(\ref{cc2t}) and (\ref{cc3t}) read as
\bea
\label{c1}
\left(\frac{M_{,\tau}}{M} \right)^2 + 16 k M = 0~,\\
\label{c2}
\frac{11}{16} \left(\frac{M_{,\tau}}{M} \right)^2 - \frac{1}{2}
\frac{M_{,\tau\tau}}{M} - k M = 0~,
\eea
It seems that the parametrization (\ref{ttau}) is also a bit
awkward. In fact, by putting (\ref{Msol}) into the constraints (\ref{c1}) and
(\ref{c2}) from both of them one gets a very restrictive condition on the solution
(\ref{Msol}) such as
\bea
\label{Mcond}
\beta^2 \left[ \cosh{\beta \tau} +
\sqrt{1 - \frac{4k}{\beta^2}} \sinh{\beta \tau}\right]^2 &=& 0~.
\eea
Without the requirement of restricting the values of the cosmic
time this condition necessarily requires that the constant
\bea
\beta &=& 0~.
\eea
This, on the other hand, in the limit $\beta \to 0$ gives from (\ref{Msol}) that
\bea
\frac{1}{\sqrt{M(\tau)}} &=& 1 + 2 \sqrt{-k} \tau~,
\eea
(so that this solution holds only for $k=-1$ models) which then from (\ref{Mtau}) gives
\bea
\label{32atau}
a(\tau) &=& \frac{e^{\frac{\alpha\tau}{2}}}{(1 + 2 \sqrt{-k} \tau)^{1/2}}~,\\
\label{32phitau}
\phi(\tau) &=& \alpha \tau~.
\eea
From the form of the above solutions one can immediately see that
there exists only the solution for negative curvature $k=-1$ Friedmann
models and that the solutions for $k=+1$ is not admissible at all within
the framework of conformal cosmology.
It is advisable to notice that the equations (\ref{c1})-(\ref{c2})
are equivalent to
\bea
\label{c3}
M_{'\tau\tau} + 24 kM^2 &=& 0~,\\
\label{c4}
M_{'\tau}^2 + 16 kM^3 &=& 0~.
\eea
These equations suggest an appropriate change of time coordinate
as
\bea
d\tau &=& \frac{d\eta}{2\sqrt{M}}~,
\eea
which transfers them into an easy to integrate form
\bea
\label{c5}
M_{'\eta\eta} + 4 kM &=& 0~,\\
\label{c6}
M_{'\eta}^2 + 4 kM^2 &=& 0~.
\eea
The solution of the system (\ref{c5})-(\ref{c6}) is very straightforward and reads as
\bea
M &=& M_0 e^{\pm 2\sqrt{-k}\eta}~.
\eea
On the other hand, the Eq. (\ref{cc1t}) in terms of $\eta-$time
reads as
\bea
\left[\ln{(\sqrt{M}\phi_{'\eta})}\right]_{'\eta} &=& 0~,
\eea
which solves by
\bea
\phi(\eta) &=& \frac{c_1}{\mp 2\sqrt{-k}}e^{\mp 2\sqrt{-k}\eta} +
c_2~,
\eea
where $c_1,c_2$ are constants. Using this, one has for the scale factor
\bea
a(\eta) &=& e^{\phi/2} M^{\frac{1}{4}} =
\exp{(\frac{c_1}{\mp 2\sqrt{-k}}e^{\mp 2\sqrt{-k}\eta} + c_2)}
M^{1/4}_0 e^{\pm \frac{1}{2}\sqrt{-k}\eta}~.
\eea

In order to find the status of the solution (\ref{32atau}) and
(\ref{32phitau}) in the Jordan frame, i.e., the solution of the Brans-Dicke theory
with $\omega=-3/2$, we now use the definition
of the conformal factor (\ref{PhitildetoPhi}) in terms of the
fields $\phi$ and $\tilde{\phi}$ as follows
\bea
\Omega &=& \frac{e^{-\phi/2}}{e^{-\tilde{\phi}/2}}~,
\eea
and make an appropriate transformation into the Einstein frame
in which the scalar field is minimally coupled to gravity. This
can be achieved by the assumption that one of the scalar fields is
constant. Let us assume that
\bea
\tilde{\phi} &=& \tilde{\phi}_0 = {\rm const.}
\eea
is such a field, which means that the conformal transformation from
the conformal (or Jordan/string) frame to the Einstein frame
reads as
\bea
\Omega_E &=& e^{\tilde{\phi}_0/2} e^{-\phi/2}~,
\eea
and all the quantities in the Einstein frame will then be labeled
by tildas. The conformal transformation then applied to
(\ref{cf1}) and (\ref{cf2}) with the help of (\ref{ttau}) gives
\bea
d\tilde{t} &=& a^3 e^{-\frac{3}{2} \phi} e^{\tilde{\phi}_0/2}
d\tau~,
\eea
which, after the application of (\ref{32atau}) and
(\ref{32phitau}) reads as
\bea
d\tilde{t} &=& \frac{e^{\tilde{\phi}_0/2}}{\left(1 + 2 \sqrt{-k}
\tau \right)^{3/2}} d\tau~.
\eea
Then, it produces the relation between the Einstein frame time $\tilde{t}$
and the $\tau-$time as
\bea
\label{tetau}
\tilde{t} - \tilde{t}_0 &=& -
\frac{e^{\tilde{\phi}_0/2}}{\sqrt{-k} \sqrt{1 + 2\sqrt{-k}
\tau}}~,
\eea
where $\tilde{t}_0=$ const. Using (\ref{cf2}) and (\ref{tetau}) we
have in the Einstein frame
\bea
\tilde{a}(\tilde{t}) &=& \sqrt{-k} (\tilde{t}_0 - \tilde{t})~.
\eea
Due to the sign choice freedom in (\ref{cf1})-(\ref{cf2}) and
taking without the loss of generality $\tilde{t}_0=0$ one can
write down this solution as
\bea
\tilde{a}(\tilde{t}) &=& \sqrt{-k} \mid \tilde{t} \mid~.
\eea
This, on the other hand, is just the Milne model which is
equivalent to Minkowski space. In order to check whether it is consistent
let us just study this problem starting directly from the Brans-Dicke theory
in the Eintein frame.

The Brans-Dicke action in the Einstein frame reads as (see e.g.
\cite{relatnn})
\bea
\label{Eframe}
\tilde{S} &=& \frac{1}{16\pi} \int d^4x \sqrt{-\tilde{g}} \left[
 \tilde{R}  - \left( \om + \frac{3}{2} \right)
\stackrel{\sim}{\partial}_{\mu}\phi\stackrel{\sim}{\partial}^{\mu}\phi
 \right]~
\eea
which for $\om=-3/2$ gives exactly the Einstein-Hilbert action
(no matter energy momentum tensor). The resulting Einstein frame
equations for (\ref{FRWtildemetric}) are \cite{relatnn}
\bea
\label{Efr1}
\left( \om + \frac{3}{2} \right) \left[\ddot{\phi} + 3 \frac{\tid}{\tia}\dot{\phi} \right] &=& 0~,\\
\label{Efr2}
3 \frac{{\tid}^2 + k}{\tia^2} &=&  \left( \om + \frac{3}{2} \right)
\dot{\phi}^2~,\\
\label{Efr3}
- 2 \frac{\tidd}{\tia} - \frac{{\tid}^2 + k}{\tia^2} &=&
\left( \om + \frac{3}{2} \right) \dot{\phi}^2~,
\eea
where the dot in these equations represents a differentiation with
respect to $\tilde{t}$.
The solutions of (\ref{Efr1})-(\ref{Efr3}) for an arbitrary value
of the parameter $\om \neq - 3/2$ and $k=0$ read as
\bea
\tia &=& \mid \tit \mid^{\frac{1}{3}}~,\\
\phi &=& \phi_0 + \frac{1}{\sqrt{3(\om+\frac{3}{2})}} \ln{\mid
\tit \mid}~.
\eea
First notice that for $\om = -3/2$ and $k=0$ the unique solution gives
\bea
\label{tid0}
\tid &=& 0~, \phi - {\rm arbitrary}~,
\eea
which is just a flat Minkowski universe. This claim seems to be consistent with
our solution (\ref{cgrs1}) in the Jordan frame. Finally, for the case of our interest,
$k \neq 0$, we get from (\ref{Efr2}) that
\bea
\tia &=& \sqrt{-k} \hspace{5pt} \mid \tit \mid~, \phi - {\rm
arbitrary}~,
\eea
which is admissible only for $k=-1$ and this solution represents
Milne universe \cite{narlikar} (in which there is no acceleration
of the expansion since the deceleration parameter
$q = \ddot{\tilde{a}}\tilde{a}/\dot{\tilde{a}}^2=0$).
However, its relation to Minkowski spacetime
\bea
dS^2 &=& - dT^2 + dR^2 + R^2 d\theta^2 + R^2 \sin^2{\theta} d\varphi^2
\eea
requires coordinate transformation
\bea
T &=& \tilde{t} \sqrt{1+r^2}, \hspace{0.5cm} R = \tit r~,
\eea
which involves the two time scales - a dynamical
one $\tit$ and an atomic one $T$ \cite{narlikar,bbpp} which may be responsible for
the cosmological redshift effect. On the other hand, the solution for $k=+1$
would be possible only if the cosmological constant was admitted -
again, cosmological redshift in this Static Einstein model would
be the result of a different time scaling \cite{narlikar}. One
should also notice that it is easy to add the cosmological
constant term into the action (\ref{Eframe}) (which is equivalent
to self-interaction potential with $\tilde{\Phi}=$ const. in
(\ref{selfinteraction})). This would allow for a non-flat Anti-deSitter (or a deSitter)
solution as in Ref. \cite{jackiw}, which would then be transformed
into the Jordan frame with $\Phi \neq$ const.

In conclusion, it seems that the reason for having only the flat solutions
in $\omega=-3/2$ Brans-Dicke cosmology is
that in the Einstein frame action (\ref{Eframe}) in this limit, the
kinetic term of the scalar field vanishes, and the action is
equivalent to a vacuum Einstein-Hilbert action, which necessarily
admits only vacuum (i.e. flat) solutions.

On the other hand, the time scaling of the scale factor for the
Milne model is the same as the scaling for the cosmological fluid
of cosmic strings $p=-(1/3)\varrho$ which have negative pressure.
This fact seems to be consistent with the supernovae data which
requires negative pressure for having cosmic acceleration, although
it is not strong enough to be fully consistent with phantom $p<-\varrho$
matter, which is favoured with the most recent data \cite{riess2004}.

%Note that the solutions in a conformally related frame with tildes
%are the same due to the relation (\ref{PhitildetoPhi}).

\section{Conformal Kantowski-Sachs cosmology}

\setcounter{equation}{0}

In this Section we choose an anisotropic Kantowski-Sachs form
of the metric of spacetime, with \cite{kantowski}
\begin{equation}
ds^2=-dt^2+X^2(t)dr^2+Y^2(t)d\Omega_k^2,
\end{equation}
where the angular metric is
\begin{equation}
d\Omega_k^2=d\theta^2+S^2(\theta)d\psi^2,
\end{equation}
and
\begin{equation}
S(\theta) = \left\{ \begin{array}{lll}
\sin(\theta) & \textrm{for}& k=+1,\\
\theta & \textrm{for} & k=0,\\
\sinh(\theta) & \textrm{for} &k=-1.
\end{array} \right.
\end{equation}
The functions $X(t)$ and $Y(t)$ are the expansion scale factors. We shall consider
models of all three curvatures in the same analysis, although the
pure Kantowski-Sachs models are of $k=+1$ spatial curvature. For $k=0$
we deal with axisymmetric Bianchi type I models, while for $k=-1$ we
deal with Bianchi III models.

After the conformal transformation (\ref{conf_trafo}) one can see that
the transformed metric is
\begin{equation}
d\tilde{s}^2=-d\tilde{t}^2+\tilde{X}^2(\tilde{t})dr^2+\tilde{Y}^2(\tilde{t})d\Omega_k^2,
\end{equation}
and the time coordinates and scale factors are related via \cite{clw1,bd97}
\bea
d\tilde{t} &=& \Omega dt~,\\
\tilde{X} &=& \Omega X~,\\
\tilde{Y} &=& \Omega Y~.
\eea

\noindent The nonzero Ricci tensor components are \cite{bd97}
\begin{equation}
R^0_0=\frac{\ddot{X}}{X}+2\frac{\ddot{Y}}{Y},
\end{equation}
\begin{equation}
R^1_1=\frac{\ddot{X}}{X}+2\frac{\dot{Y}}{Y}\frac{\dot{X}}{X},
\end{equation}
\begin{equation}
R^2_2=R^3_3=\frac{k+\dot{Y}^2}{Y^2}+\frac{\ddot{Y}}{Y}+\frac{\dot{Y}}{Y}\frac{\dot{X}}{X},
\end{equation}
and the scalar curvature is
\begin{equation}
R=2\frac{\ddot{X}}{X}+4\frac{\ddot{Y}}{Y}+2\frac{k+\dot{Y}^2}{Y^2}+4\frac{\dot{Y}}{Y}\frac{\ddot{X}}{X}.
\end{equation}
The field equations (\ref{pbb2}) become
\begin{equation}
\frac{\ddot{X}}{X}+2\frac{\ddot{Y}}{Y}-\ddot{\phi}-(\omega+1)\left(-\frac{\dot{X}}{X}\dot{\phi}-2\frac{\dot{Y}}{Y}\dot{\phi}-\ddot{\phi}\right)=0,\label{B}
\end{equation}
\begin{equation}
\frac{\ddot{X}}{X}+2\frac{\dot{Y}}{Y}\frac{\dot{X}}{X}-\frac{\dot{X}}{X}\dot{\phi}-(\omega+1)\left(\dot{\phi}^2-\frac{\dot{X}}{X}\dot{\phi}-2\frac{\dot{Y}}{Y}\dot{\phi}-\ddot{\phi}\right)=0,\label{C}
\end{equation}
\begin{equation}
\frac{k+\dot{Y}^2}{Y^2}+\frac{\ddot{Y}}{Y}+\frac{\dot{Y}}{Y}\frac{\dot{X}}{X}-\frac{\dot{Y}}{Y}\dot{\phi}-(\omega+1)\left(\dot{\phi}^2-\frac{\dot{X}}{X}\dot{\phi}-2\frac{\dot{Y}}{Y}\dot{\phi}-\ddot{\phi}\right)=0\label{D}
\end{equation}
The field equation (\ref{pbb1}) reads
\begin{equation}
2\frac{\ddot{X}}{X}+4\frac{\ddot{Y}}{Y}+2\frac{k+\dot{Y}^2}{Y^2}+4\frac{\dot{Y}}{Y}\frac{\dot{X}}{X}-\omega\dot{\phi}^2+2\omega\ddot{\phi}+2\omega\left(\frac{\dot{X}}{X}+2\frac{\dot{Y}}{Y}\right)\dot{\phi}=0.\label{A}
\end{equation}
Adding the Eqs. (\ref{B}), (\ref{C}) with doubled (\ref{D}) and
subtracting from this sum Eq. (\ref{A}) we get
\begin{equation}
\ddot{\phi}-\dot{\phi}^2+\left(\frac{\dot{X}}{X}+2\frac{\dot{Y}}{Y}\right)\dot{\phi}=0.\label{2.21}
\end{equation}
At this stage we introduce a new time coordinate $\tau$ via
relation
\begin{equation}
dt=XY^2e^{-\phi}d\tau.\label{time}
\end{equation}
Then Eq. (\ref{2.21}) becomes
\begin{equation}
\phi_{,\tau\tau}=0~,
\end{equation}
which solves as
\begin{equation}
\phi(\tau)=a\tau+\gamma.\label{fi od tau}
\end{equation}
Using the time coordinate (\ref{time}) equations
(\ref{B})--(\ref{D}) become
\begin{equation}
\left(\frac{X_{,\tau}}{X}\right)_{,{\tau}}+2\left(\frac{Y_{,\tau}}{Y}\right)_{,{\tau}}-2\frac{Y_{,\tau}}{Y}\left(\frac{Y_{,\tau}}{Y}+2\frac{X_{,\tau}}{X}\right)+2\phi_{,\tau}\left(\frac{X_{,\tau}}{X}+2\frac{Y_{,\tau}}{Y}\right)+\omega\phi^2_{,\tau}=0,\label{B_od_tau}
\end{equation}
\begin{equation}
\left(\frac{X_{,\tau}}{X}\right)_{,{\tau}}=0,\label{C_od_tau}
\end{equation}
\begin{equation}
\left(\frac{Y_{,\tau}}{Y}\right)_{,{\tau}}+kX^2Y^2e^{-2\phi}=0.\label{D_od_tau}
\end{equation}
The solution of equation (\ref{C_od_tau}) is simply
\begin{equation}
X=\frac{1}{A_0}e^{c\tau},\label{x od tau}
\end{equation}
where $c$ and $A_0$ are a constants. When we put in the Eq.
(\ref{x od tau}) $A_0=1$ and $c=\frac{1}{2}(a+p)$, where $a$ is
taken from (\ref{fi od tau}) and $p$ is a constant, we obtain
(\ref{x od tau}) in the following form
\begin{equation}
X(\tau)=e^{\frac{1}{2}(a+p)\tau}
\end{equation}
In Eq. (\ref{fi od tau}) we set $\gamma=0$ without loss of
generality. Then from Eqs. (\ref{B_od_tau}) and (\ref{D_od_tau})
we obtain
\begin{equation}
Y(\tau)=\left\{ \begin{array}{lll}
e^{\frac{1}{2}(a-p)\tau}\sqrt{\frac{1}{4}[(2\omega+3)a^2+p^2]}\{\cosh[\frac{1}{4}(a^2\omega+3a^2+p^2)]\tau\}^{-1/2} & \textrm{for}& k=+1,\\
e^{\frac{1}{2}(a-p)\tau}e^{-\sqrt{\frac{1}{4}[(2\omega+3)a^2+p^2]}\tau} & \textrm{for} & k=0,\\
e^{\frac{1}{2}(a-p)\tau}\sqrt{\frac{1}{4}[(2\omega+3)a^2+p^2)}\{\sinh[\frac{1}{4}(a^2\omega+3a^2+p^2)]\tau\}^{-1/2}
& \textrm{for} &k=-1,
\end{array}\right.
\end{equation}
For $k=0$ after integration of Eq. (\ref{time}) we get
\begin{equation}
t(\tau)=\frac{-2e^{-\frac{1}{2}\tau\left(p-a+2\sqrt{p^2+a^2(3+2\omega)}\right)}}{\left(-a+p+2\sqrt{p^2+a^2(2+3\omega)}\right)}.
\end{equation}
And for $k\neq0$ Eq. (\ref{time}) is integrable for a=p. We get
\begin{equation}
t(\tau)=\left\{ \begin{array}{lll}
\pm\frac{a}{\sqrt{2}}\sqrt{2+\omega}\coth\left[\pm\frac{a}{\sqrt{2}}\sqrt{2+\omega}\tau\right] & \textrm{for}& k=-1,\\
\mp\frac{a}{\sqrt{2}}\sqrt{2+\omega}\coth\left[\mp\frac{a}{\sqrt{2}}\sqrt{2+\omega}\tau\right]
& \textrm{for} & k=+1.
\end{array} \right.
\end{equation}
Then for $k=0$ (axisymmetric Bianchi I type) we obtain solution in the form
\begin{equation}
\label{Xk0}
X(t)=\left\{-\frac{1}{2}t\left(-a+p+2\sqrt{p^2+a^2(3+2\omega)}\right)\right\}^{-\frac{a+p}{p-a+2\sqrt{p^2+a^2(3+2\omega)}}},
\end{equation}
\begin{equation}
\label{Yk0}
Y(t)=\left\{-\frac{1}{2}t\left(-a+p+2\sqrt{p^2+a^2(3+2\omega)}\right)\right\}^{\frac{p-a+\sqrt{p^2+a^2(3+2\omega)}}{p-a+2\sqrt{p^2+a^2(3+2\omega)}}},
\end{equation}
\begin{equation}
\label{phik0}
\phi(t)=\frac{2a}{a-p-2\sqrt{p^2+a^2(3+2\omega)}}\ln\left\{-\frac{1}{2}t\left(p-a+2\sqrt{p^2+a^2(3+2\omega)}\right)\right\}.
\end{equation}
These solutions generalize the isotropic solution given by
(\ref{cgrs1}) and (\ref{cgrs2}). Taking $\om=-3/2$ we get
\begin{equation}
\label{Xk032}
X(t)=\left\{\frac{1}{2}t\left(a-3p \right)\right\}^{\frac{a+p}{a-3p}},
\end{equation}
\begin{equation}
\label{Yk032}
Y(t)=\left\{\frac{1}{2}t\left(a-3p \right)\right\}^{\frac{a-2p}{a-3p}},
\end{equation}
\begin{equation}
\label{phik032}
\phi(t)=\frac{2a}{a-3p}\ln\left\{\frac{1}{2}t\left(a-3p \right)\right\}.
\end{equation}

\begin{figure}[h!]\begin{center}
\includegraphics[width=12cm]{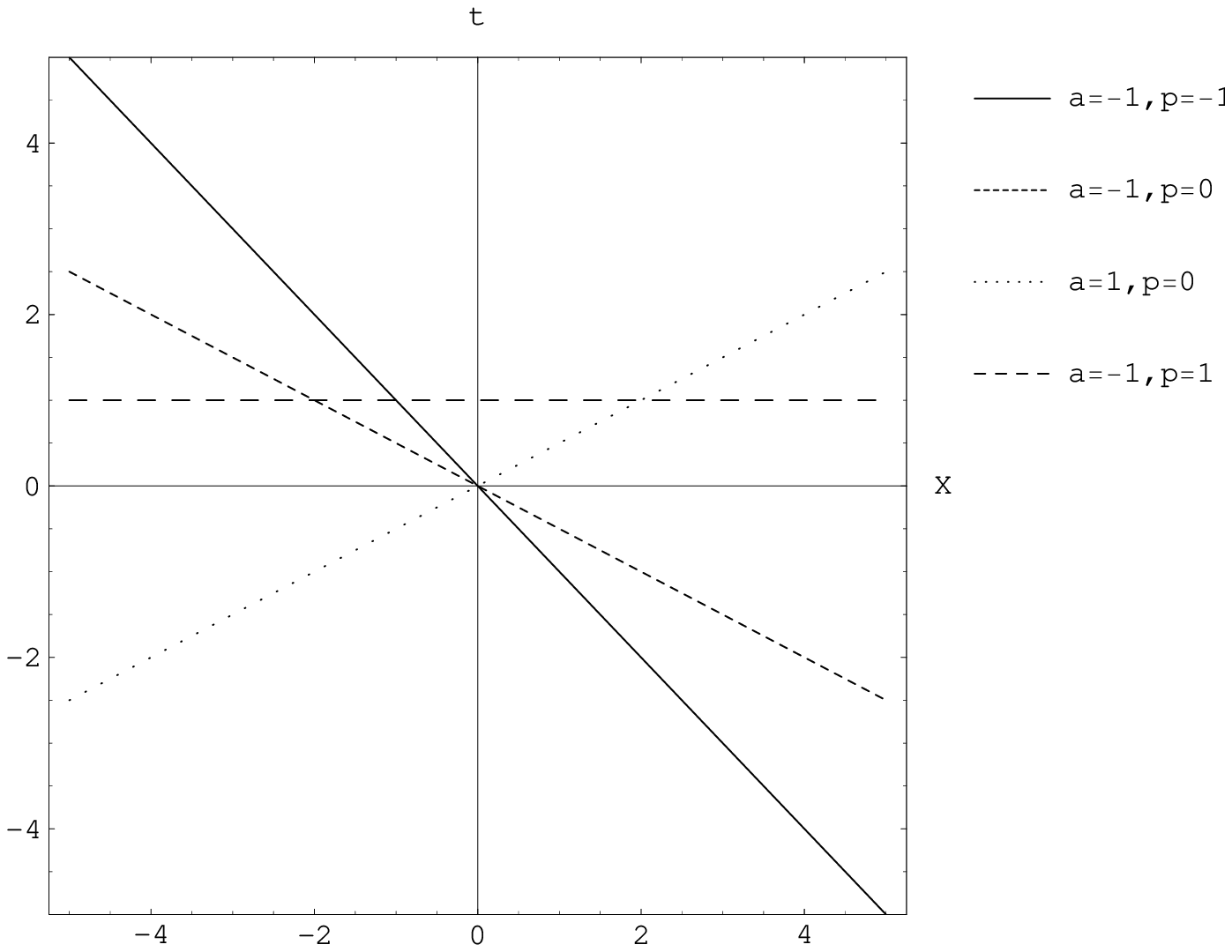}
\caption{The plots of the scale factor $X$ (Eq. (\ref{Xk0})) in conformal relativity
$(\omega=-3/2)$ for the axisymmetric Bianchi I $(k=0)$
cosmological models. Different shapes of the plots depend on the values of the
constants $a$ and $p$.}
\label{fig01}\end{center}
\end{figure}

\begin{figure}[h!]\begin{center}
\includegraphics[width=12cm]{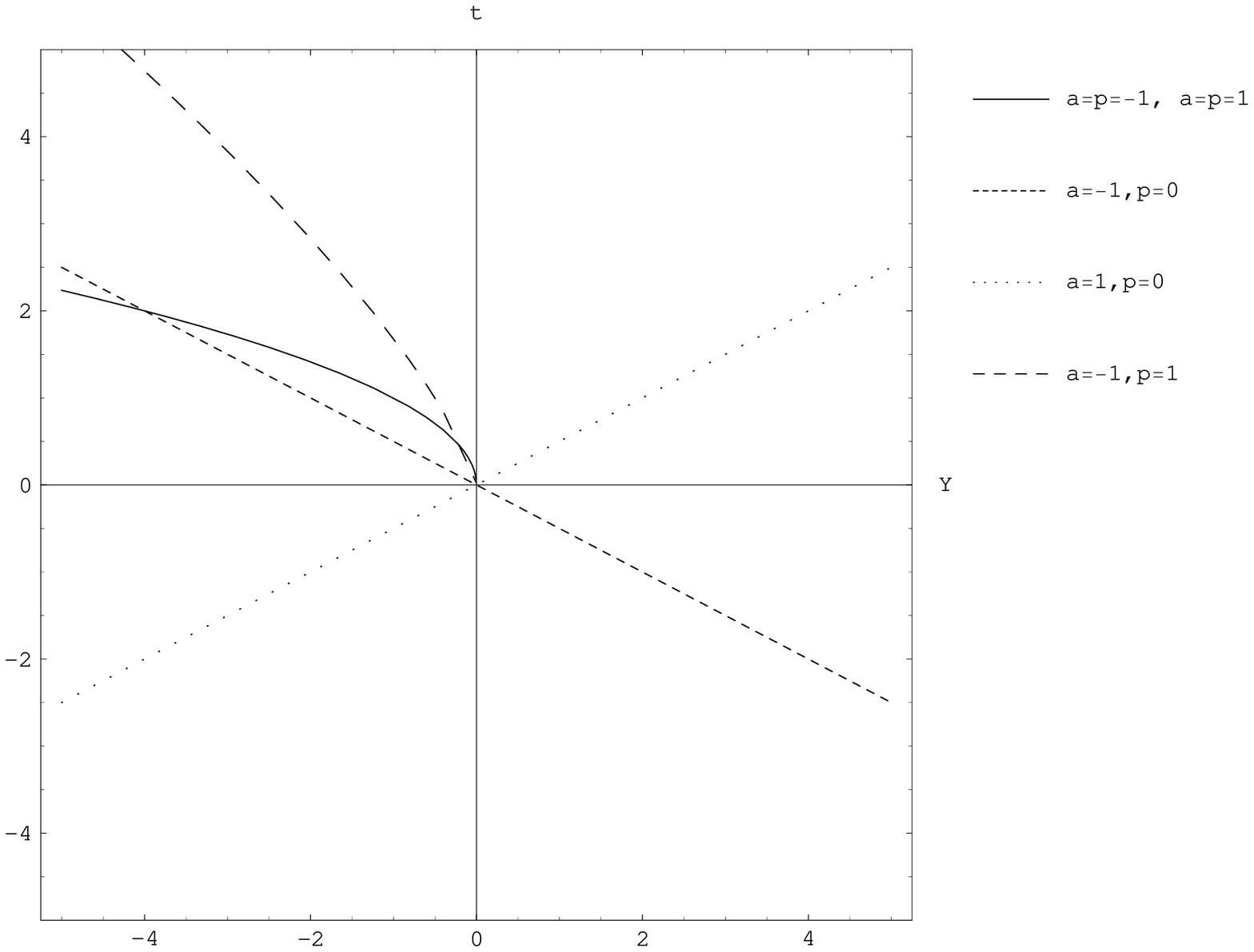}
\caption{The plots of the scale factor $Y$ (Eq. (\ref{Yk0})) in conformal relativity
$(\omega=-3/2)$ for the axisymmetric Bianchi I $(k=0)$
cosmological models. Different shapes of the plots depend on the values of the
constants $a$ and $p$.}
\label{fig02}\end{center}
\end{figure}

\begin{figure}[h!]\begin{center}
\includegraphics[width=12cm]{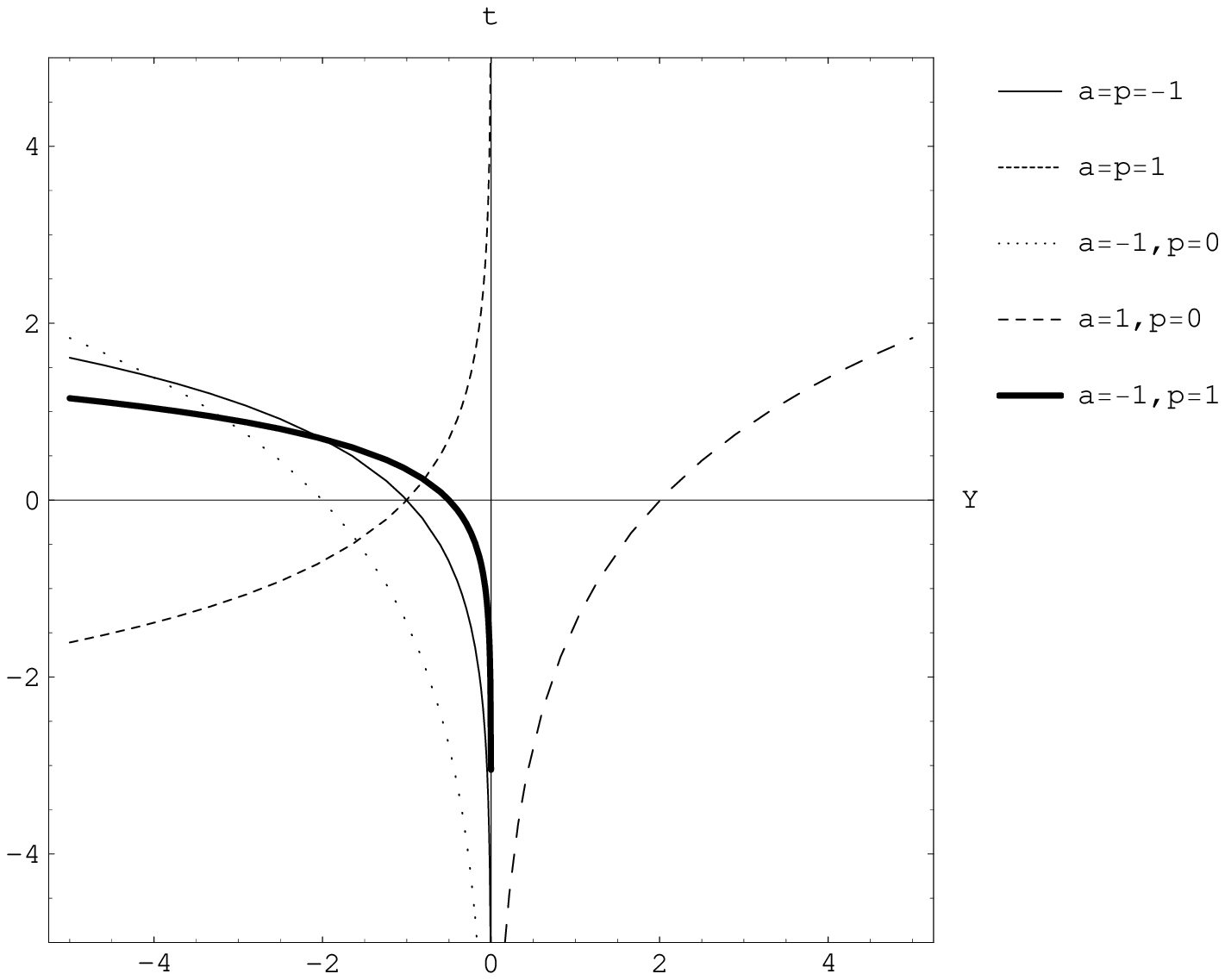}
\caption{The plots of the dilaton field $\phi$ (Eq. (\ref{phik0})) in conformal relativity
$(\omega=-3/2)$ for the axisymmetric Bianchi I $(k=0)$
cosmological models. Different shapes of the plots depend on the values of the
constants $a$ and $p$.}
\label{fig03}\end{center}
\end{figure}

The plots of these solutions for
conformal relativity $(\om=-3/2)$ and for the different values of the parameters
$a$ and $p$ are given in Figs. \ref{fig01}, \ref{fig02}, and \ref{fig03}.

\begin{figure}[h!]\begin{center}
\includegraphics[width=12cm]{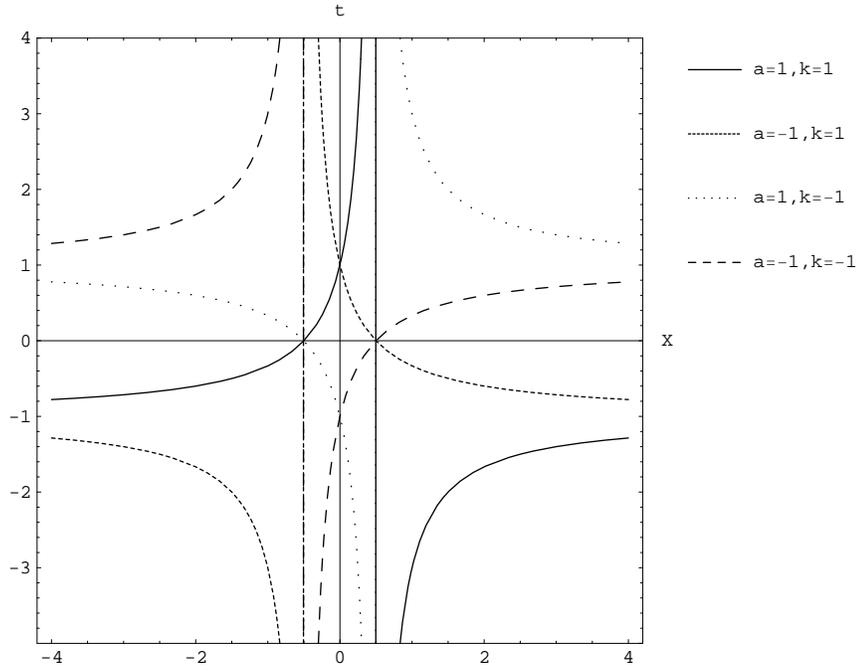}
\caption{The plots of the scale factor $X$ (Eq. (\ref{Xk1})) in conformal relativity
$(\omega=-3/2)$ for Kantowski-Sachs $(k=+1)$ and Bianchi III $(k=-1)$
cosmological models. Different shapes of the plots depend on the values of the constant $a$.}
\label{fig1}\end{center}
\end{figure}

\begin{figure}[h!]\begin{center}
\includegraphics[width=12cm]{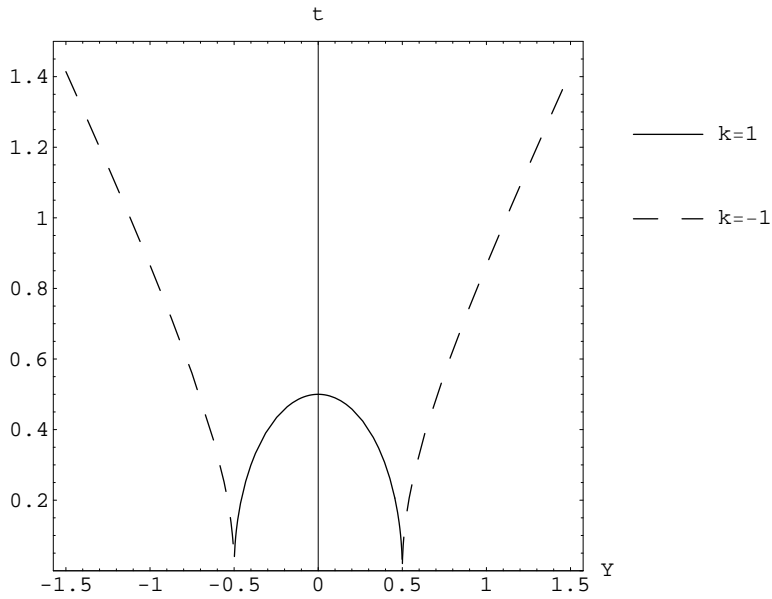}
\caption{The plots of the scale factor $Y$ (Eq. (\ref{Yk1})) in conformal relativity
$(\omega=-3/2)$ for Kantowski-Sachs $(k=+1)$ and Bianchi III $(k=-1)$
cosmological models.}
\label{fig2}\end{center}
\end{figure}

\begin{figure}[h!]\begin{center}
\includegraphics[width=12cm]{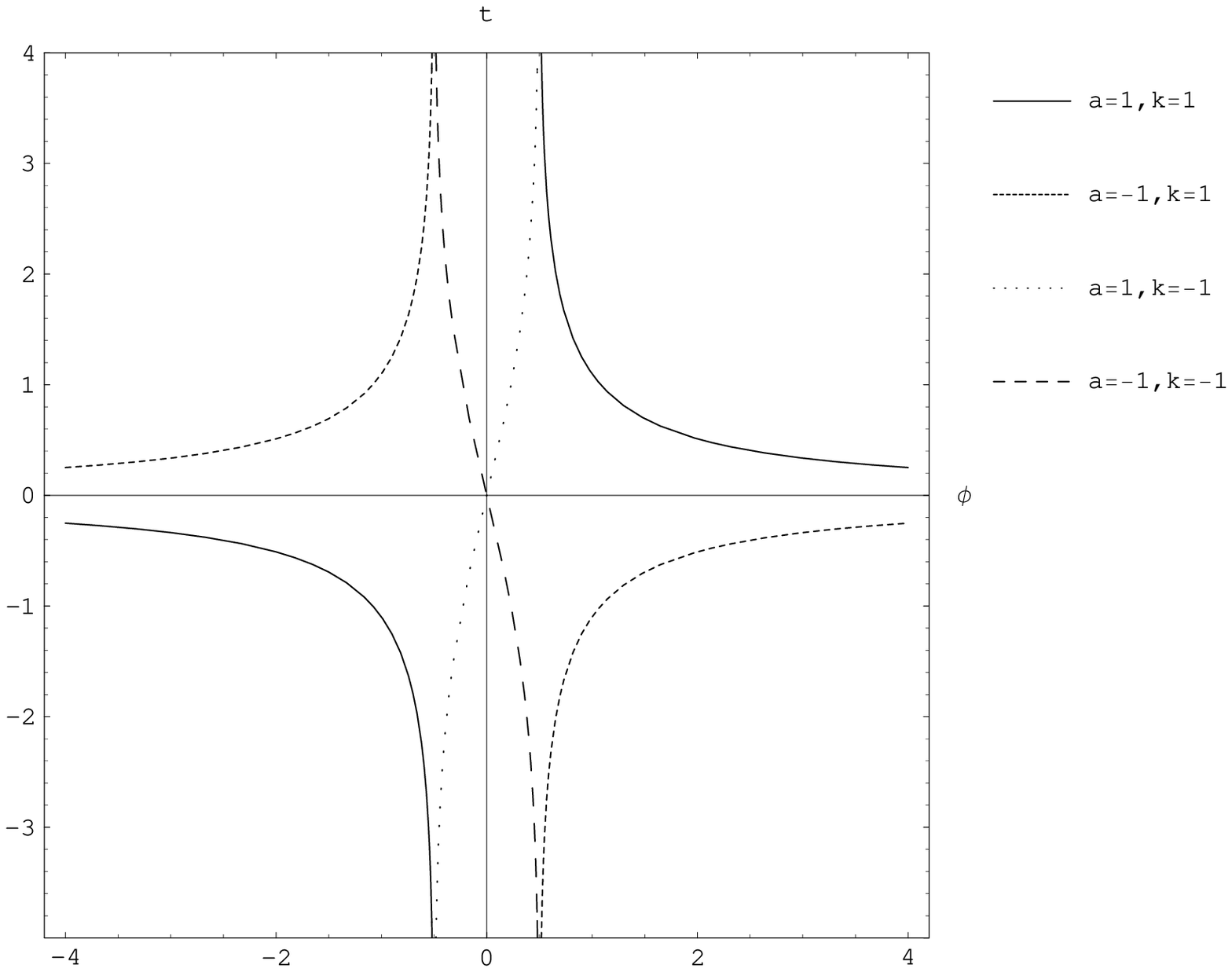}
\caption{The plots of the dilaton field $\phi$ (Eq. (\ref{phik1})) in conformal relativity
$(\omega=-3/2)$ for Kantowski-Sachs $(k=+1)$ and Bianchi III $(k=-1)$
cosmological models. Different shapes of the plots depend on the values of the constant $a$.}
\label{fig3}\end{center}
\end{figure}

For non-zero $k=\pm1$ curvature we have
\begin{equation}
\label{Xk1}
X(t)=\left\{k\frac{\frac{a\sqrt{2+\omega}}{\sqrt{2}}+t}
{\frac{a\sqrt{2+\omega}}{\sqrt{2}}-t}\right\}^{\frac{1}{\sqrt{2(2+\omega)}}}
\end{equation}
\begin{equation}
\label{Yk1}
Y(t)=\sqrt{k\left[\frac{a^2(2+\omega)}{2}-t^2\right]}
\end{equation}
\begin{equation}
\label{phik1}
\phi(t)=\ln\left\{k\frac{t+\frac{a\sqrt{2+\omega}}{\sqrt{2}}}
{t-\frac{a\sqrt{2+\omega}}{\sqrt{2}}}\right\}^{\frac{1}{\sqrt{2(2+\omega)}}}
\end{equation}
In order to understand the nature of both initial and final
singularities it is important to study the evolution of the volume
\begin{equation}
\label{Vol1}
V(t) = X(t)Y^2(t) = \left[ k \left( \frac{a\sqrt{2+\omega}}{\sqrt{2}} + t\right) \right]
^{1 + \frac{1}{\sqrt{2(2+\omega)}}}
\left[ k \left( \frac{a\sqrt{2+\omega}}{\sqrt{2}} - t\right) \right]
^{1 - \frac{1}{\sqrt{2(2+\omega)}}}~.
\end{equation}
In the case of conformal relativity $\omega=-3/2$ we have
\bea
X &=& k \frac{\frac{a}{2}+t}{\frac{a}{2}-t}~,\\
Y^2 &=& k \left( \frac{a}{2}+t \right) \left( \frac{a}{2}-t
\right)~,\\
\phi &=& \ln{\left[k \frac{t+\frac{a}{2}}{t-\frac{a}{2}} \right]}~,
\eea
where the time coordinate has the ranges
\bea
0 &\leq& t^2 \leq \frac{a^2}{4} \hspace{0.2cm} {\rm for} \hspace{0.2cm} k=+1~,\\
t^2 &\geq& \frac{a^2}{4} \hspace{0.2cm} {\rm for} \hspace{0.2cm}
k=-1~.
\eea
The volume (\ref{Vol1}) scales as
\begin{equation}
V = \left( \frac{a}{2} + t \right)^2~,
\end{equation}
which shows that the divergent term for $t=a/2$ was cancelled. This means
we deal with initial Big-Bang type of singularity at $t = -a/2$
where $X=Y=0$ (though it is weak coupling since $e^{\phi} \to 0$)
while at $t=a/2$ the volume is finite despite the
fact that $X \to \infty$ and $Y=0$ there and suggests the appearance of the
barrel singularity (though it is strong coupling since $e^{\phi} \to \infty$).
The plots of the solutions (\ref{Xk1}) and (\ref{Yk1}) for
$\omega=-3/2$ are given in Figs. \ref{fig1}, \ref{fig2}, and \ref{fig3}. The
string cosmology case $\omega=-1$ was given in Ref. \cite{bd97}.

\vspace{0.3cm}

\section{Conclusions}

We have studied isotropic Friedmann, anisotropic
Kantowski-Sachs, axisymmetric Bianchi I, and Bianchi III
cosmological models within the framework of $\omega=-3/2$ Brans-Dicke
cosmology which is equivalent to the so-called conformal relativity.
In fact, we have started from a general class of solutions which are
the vacuum Brans-Dicke theory solutions in the Jordan frame. These solutions
are parametrized by the Brans-Dicke parameter $\omega$. The
conformal relativity solutions are given for $\om=-3/2$, while the
low-energy-superstring (pre-big-bang) type of solutions are given for
$\om=-1$. It emerged that the conformal relativity limit $\om=-3/2$ is
nontrivial and cannot be obtained automatically from the
Brans-Dicke solutions by most of the routine Brans-Dicke time parametrizations.
Despite that, we were able to find an appropriate time
parametrization to show that there exist only the $k=0$ and $k=-1$
isotropic solutions in the Jordan frame. We have also shown that
the $k=-1$ solution in the Jordan frame represents the Milne
universe in the conformally related (with scalar field minimally coupled to gravity)
Einstein frame which, in turn, is equivalent to a flat Minkowski
spacetime. Because of that, we claim that only the flat models are
consistent with $\omega=-3/2$ Brans-Dicke cosmology. In a way this
is not a surprise since in the Einstein frame the kinetic term of
the scalar field vanishes for $\omega=-3/2$.

An additional point of interest in the $\omega=-3/2$ solutions is
the fact that the recent fit to supernovae data \cite{supernovae} shows, that despite
local gravitational tests which give the constraint $\omega > 1000$,
supernovae favour exactly the value of $\omega = - 3/2$ \cite{fabris}.
Besides, $\omega = -3/2$ gives a border line between
a standard scalar field model and a ghost/phantom model in the Einstein frame \cite{bd,phantom}.
This may be one of the crucial points for the success of the fit
in the Jordan frame although the time scaling of the scale factor for the
Milne universe is the same as the scaling for the cosmological fluid
of cosmic strings $p=-(1/3)\varrho$ in the Einstein frame which is not strong enough to be
fully consistent with phantom $p<-\varrho$
matter favoured by the most recent supernovae data \cite{riess2004}.

Apart from isotropic solutions we have also studied
anisotropic Kantowski-Sachs, axisymmetric Bianchi I, and Bianchi III type solutions.
In particular, anisotropic Kantowski-Sachs models
of non-zero spatial curvature are admissible in $\omega=-3/2$ Brans-Dicke theory, i.e.,
in conformal relativity. This means that an additional scale factor which
appears in Kantowski-Sachs models gives an extra degree of freedom
to the theory and makes it less restrictive than in an isotropic
Friedmann case although these solutions should be conformally equivalent to
vacuum solutions in the Einstein frame. In our paper these anisotropic solutions were fully
deparametrized in terms of the cosmic time $t$ and not given in
terms of the parametric time only as in the previous literature.

Besides, in the isotropic Friedmann case, the advantage of conformal
relativity solutions to pre-big-bang solutions is that there is no
strong coupling singularity accompanied to a curvature (Big-Bang) singularity
for these models. This is in analogy to ekpyrotic models which
have intensively been studied recently.

However, in the anisotropic case, the problem of transition
through singularity at weak coupling regime is more complicated
and depends on the parameters of the models. This makes some
motivation to study such anisotropic models and possibly also some
inhomogeneous models within the framework of $\omega=-3/2$ Brans-Dicke cosmology.

\newpage

\section*{Acknowledgments}

The authors thank John Barrow and David Polarski for useful comments.
D.B. and T.D. have been supported by the DFG Graduiertenkolleg No. 567
{\it Strongly Correlated Many-Particle Systems} at the University of
Rostock. M.P.D. acknowledges the support from the Polish Research Committee (KBN)
grant No 2P03B 090 23. M.P.D. and T.D. acknowledge the support from the Ministry of
Education and Science grant No 1P03B 043 29 (years 2005-07).

\appendix

\section{Conformal transformations}

The determinant of the metric $g={\rm det}~g_{\mu\nu}$ transforms as
\bea
\label{det}
\sqrt{-\tilde{g}} &=& \Omega^4 \sqrt{-g}~.
\eea
It is obvious from (\ref{conf_trafo}) that the following relations
for the inverse metrics and the spacetime intervals hold
\bea
\label{conf_trafo_inv}
\tilde{g}^{\mu\nu} &=& \Omega^{-2} g^{\mu\nu}~, \\
\d \tilde{s}^2 &=& \Omega^2 \d s^2~.
\eea

The application of (\ref{conf_trafo}) to the Christoffel connection coefficients
gives \cite{hawk_ellis}
\bea
\label{connections}
\tilde{\Ga}^{\la}_{\mu\nu}
&=&
\Ga^{\la}_{\mu\nu} + \frac{1}{\Om}\left( g^{\la}_{\mu} \Om_{,\nu} +
  g^{\la}_{\nu} \Om_{,\mu} - g_{\mu\nu}g^{\la\ka}\Om_{,\ka} \right)~,
\\
\label{connections1}
\Ga^{\la}_{\mu\nu}
&=&
\tilde{\Ga}^{\la}_{\mu\nu}- \frac{1}{\Om} \left( \tilde{g}^{\la}_{\mu}
  \Om_{,\nu} + \tilde{g}^{\la}_{\nu} \Om_{,\mu} -
  \tilde{g}_{\mu\nu}\tilde{g}^{\la\ka}\Om_{,\ka} \right)~.
\eea

The Ricci tensors and Ricci scalars in the two related frames $g_{\mu\nu}$ and
$\tilde{g}_{\mu\nu}$ transform as
\bea
\label{riccitensor1}
\tilde{R}_{\mu\nu}
&=&
R_{\mu\nu} + \Om^{-2}\left [
  4\Om_{,\mu}\Om_{,\nu}-\Om_{,\si}\Om^{,\si}g_{\mu\nu}\right ]
-\Om^{-1}\left [ 2\Om_{;\mu\nu}+\Box \Om g_{\mu\nu} \right ]~,
\\
\label{riccitensor2}
R_{\mu\nu} &=& \tilde{R}_{\mu\nu} - 3\Om^{-2} \Om_{,\rho}\Om^{,\rho}\tilde{g}_{\mu\nu}
+\Om^{-1}\left[2\Om_{\tilde{;}\mu\nu}+ \tilde{g}_{\mu\nu} \stackrel{\sim}{\Box} \Om  \right ]~,
\eea
\bea
\label{ricciscalar4}
\tilde{R} &=& \Om^{-2} \left [ R - 6\frac{\Box{\Om}}{\Om}
\right]~,\\
\label{ricciscalar5}
R &=& \Omega^2 \left[ \tilde{R} + 6
\frac{\stackrel{\sim}{\Box}\Om}{\Om} - 12 \tilde{g}^{\mu\nu}
\frac{\Om_{,\mu}\Om_{,\nu}}{\Om^2}\right ]~,
\eea
and the appropriate d'Alambertian operators change under (\ref{conf_trafo}) as
\bea
\stackrel{\sim}{\Box}\phi &=&\Om^{-2}\left( {\Box}\phi+
  2g^{\mu\nu}\frac{\Om_{,\mu}}{\Om}\phi_{,\nu} \right )~,\\
\Box\phi &=&\Om^{2}\left( \stackrel{\sim}{\Box}\phi-
  2\tilde{g}^{\mu\nu}\frac{\Om_{,\mu}}{\Om}\phi_{,\nu} \right )~.
\eea
In these formulas the d'Alembertian $\stackrel{\sim}{\Box}$ taken with respect to the
metric $\tilde{g}_{\mu\nu}$ is different
from $\Box$ which is taken with respect to a conformally rescaled metric
$g_{\mu\nu}$. Same refers to the covariant derivatives $\tilde{;}$
and $;$ in (\ref{riccitensor1})-(\ref{riccitensor2}).

In order to prove the conformal invariance of the field equations (\ref{eom3})
it is necessary to know the rule of the conformal
transformations for the double covariant derivative of a scalar
field, i.e.,
\bea
\label{covdertilde}
\tilde{\Phi}_{\tilde{;}\mu\nu} &=& \tilde{\Phi}_{,\mu\nu} - \tilde{\Ga}^{\rho}_{\mu\nu}
\tilde{\Phi}_{,\rho} = - \Om^{-2} \Phi \Om_{;\mu\nu} + \Om^{-1}
\Phi_{;\mu\nu} + 4 \Om^{-3} \Phi \Om_{,\mu} \Om_{,\nu} \nonumber \\
&-& 2 \Om^{-2}
\left(\Phi_{,\mu}\Om_{,\nu} + \Om_{,\mu} \Phi_{,\nu} \right) -
\Om^{-3} \Phi g_{\mu\nu} \Om_{,\rho} \Om^{,\rho} + \Om^{-2}
g_{\mu\nu} \Phi_{,\rho} \Om^{,\rho} ~,
\eea
and
\bea
\label{covder}
\Phi_{;\mu\nu} &=& \tilde{\Phi} \Om_{;\mu\nu} + \Om \Phi_{;\mu\nu}
+ \frac{2}{\Om} \tilde{\Phi} \Om_{,\mu} \Om_{,\nu} +
2 \left( \Om_{,\mu} \tilde{\Phi}_{,\nu} + \tilde{\Phi}_{,\mu}
\Om_{,\nu} \right) - \frac{1}{\Om} \tilde{\Phi} \tilde{g}_{\mu\nu}
\Om_{\rho} \Om^{\rho} - \frac{1}{\Om} \tilde{g}_{\mu\nu}
\tilde{\Phi}_{,\rho} \Om^{,\rho} = 0~.
\eea

\end{document}